\documentclass[11pt,a4paper]{article}

\pdfoutput=1

\usepackage{comment}
\usepackage{andocap}

\usepackage{physics}
\usepackage{subfig}
\usepackage{float}
\usepackage{xcolor}
\usepackage{appendix}

\usepackage{graphicx}

\newcommand{\be}{\begin{equation}}
\newcommand{\ee}{\end{equation}}
\newcommand{\bea}{\begin{eqnarray}}

\newcommand{\eea}{\end{eqnarray}}
\newcommand{\nn}{\nonumber}

\newcommand{\MP}{M_\text{P}}
\newcommand{\fNY}{f_\text{NY}}

\usepackage{tensor}
\def\tns{\tensor}
\def\cR{\mathcal{R}}

\def\a{\alpha}
\def\b{\beta}
\def\c{\chi}
\def\d{\delta}

\def\eps{\epsilon}

\def\m{\mu}

\title{\centering Natural Metric-Affine Inflation: Reloaded}

\author[a]{D. Kraiko}
\author[b]{A. Racioppi}

\affiliation[a]{Department of Physics and Astronomy, Uppsala University, Box 516, SE-751 20 Uppsala, Sweden}

\affiliation[b]{National Institute of Chemical Physics and Biophysics, R\"avala 10, 10143 Tallinn, Estonia}

\emailAdd{kraiko.dima@gmail.com}
\emailAdd{antonio.racioppi@kbfi.ee}

\abstract{We revisit natural inflation within the framework of metric-affine gravity, considering the impact of a periodic non-minimal coupling between the inflaton and the Nieh-Yan term. Such a term, alone, leads to linear inflation predictions in the strong coupling limit and cannot help to rescue the natural inflation scenario. However, once an analogous non-minimal coupling with the Ricci scalar is added, agreement with data can be easily achieved. Remarkably, the scenario remains viable even with a sub-Planckian periodicity scale and relatively small (order of one) non-minimal couplings. }
\begin{document}
\maketitle

%%%%%%%%%%%%%%%%%%%%%%%%%%%%%%%%%%%%%%%%%%%%%%%%%%%%%%%%%%%%%%%%%%%%%%%%%%%%%%%%%%%%%%%%
\section{Introduction}

The $\Lambda$CDM model, the current standard model of Big Bang cosmology, is built on the cosmological principle, which states that the Universe is homogeneous and isotropic on sufficiently large scales. Different independent observations (e.g. \cite{Planck:2018vyg,SDSS2017,2dFGRS2001}) strongly support this principle, showing that the Universe is remarkably uniform across hundreds of megaparsecs. Nevertheless, explaining the uniformity of the Universe, along with its near-flat geometry, requires a mechanism operating in the very early Universe. The most compelling solution is a brief phase of accelerated expansion~\cite{Starobinsky:1980te,Guth:1980zm,Linde:1981mu,Albrecht:1982wi}, commonly known as \emph{inflation}. Additionally, inflation provides a natural origin for the tiny primordial fluctuations that later evolved into galaxies, clusters, and the cosmic web.

A simple realization of inflation is through a scalar field, called the \emph{inflaton}, whose potential energy drives the accelerated expansion. The specific shape of the inflaton potential determines both the duration of inflation and the properties of the corresponding primordial perturbations. Unfortunately, most of the simplest minimally coupled models have already been either ruled out or strongly disfavored by recent observations \cite{Planck:2018jri,BICEP:2021xfz}. Among these, \emph{natural inflation} \cite{Freese:1990rb} stood out as a particularly elegant scenario: the inflaton is realized as a pseudo-Nambu-Goldstone boson (PNGB) with a periodic potential. The periodicity naturally keeps the potential flat, reducing the need for fine-tuning. Although theoretically appealing, the original natural inflation model is now in tension with observations, prompting numerous extensions and modifications to reconcile it with current data (e.g. \cite{Kim:2004rp,Visinelli:2011jy,Czerny:2014wza,Czerny:2014xja,Croon:2014dma,Achucarro:2015rfa,Ferreira:2018nav,Antoniadis:2018yfq,Salvio:2019wcp,Simeon:2020lkd,McDonough:2020gmn,Salvio:2021lka,Salvio:2022mld,Bostan:2022swq,Salvio:2023cry,Mukuno:2024yoa,Racioppi:2024zva,Racioppi:2025pim,Lorenzoni:2024krn,Michelotti:2024bbc,Bostan:2025vkt,Bostan:2025zdt,ZentenoGatica:2026pbf} and refs. therein).

A promising approach is to embed inflation in non-minimal metric-affine gravity (MAG).
In the standard metric formulation of gravity, the affine connection is assumed to be the Levi-Civita connection, and the metric tensor is the only dynamical gravitational degree of freedom.  On the other hand, in MAG, both the metric and the affine connection are dynamical variables and their equations of motion determine the relation between them. When the gravitational   action features only a term linear in the curvature scalar and no fermions, the two approaches lead to equivalent theories (e.g.~\cite{BeltranJimenez:2019esp,Rigouzzo:2023sbb} and refs. therein), otherwise the theories are completely different~\cite{BeltranJimenez:2019esp,Rigouzzo:2023sbb,Koivisto:2005yc,Bauer:2008zj} and produce different phenomenological predictions, as recently investigated in (e.g.~\cite{Racioppi:2017spw,Jarv:2017azx,Racioppi:2018zoy,Kannike:2018zwn,Racioppi:2019jsp,Jarv:2020qqm,Gialamas:2020snr,Racioppi:2021ynx,Racioppi:2021jai,Lillepalu:2022knx,Gialamas:2023flv,Piani:2023aof,Barker:2024ydb,Dioguardi:2021fmr,Racioppi:2022qxq,Dioguardi:2022oqu,Dioguardi:2023jwa,Kannike:2023kzt,TerenteDiaz:2023kgc,Marzo:2024pyn,Iosifidis:2025wrv,Bostan:2025zdt,Gialamas:2025kef,Dioguardi:2025vci,Dimopoulos:2025fuq,Bostan:2025vkt,Dioguardi:2025mpp,Karananas:2025xcv,Barker:2025xzd,Barker:2025rzd,Barker:2025fgo,Racioppi:2025igu,Dimopoulos:2026iwq,DiBenedetto:2026onj} and refs. therein). Moreover, MAG permits not only one, but two independent two-derivative curvature invariants: the usual Ricci-like scalar and the Holst invariant~\cite{Hojman:1980kv,Nelson:1980ph,Holst:1995pc}, which can be used to construct new models~(e.g.~\cite{Hecht:1996np,BeltranJimenez:2019hrm,Langvik:2020nrs,Rigouzzo:2022yan,Shaposhnikov:2020gts,Pradisi:2022nmh,Salvio:2022suk,Piani:2022gon,DiMarco:2023ncs,Gialamas:2022xtt,Gialamas:2024jeb,Gialamas:2024iyu,Racioppi:2024zva,Racioppi:2025pim,Racioppi:2024pno,Gialamas:2024uar,He:2024wqv,He:2025bli,He:2025fij,Katsoulas:2025srh,Racioppi:2025igu,Dimopoulos:2026iwq,DiBenedetto:2026onj,Gialamas:2026pjo} and refs. therein). Natural inflation in presence of periodic non-minimal couplings with the Ricci scalar and the Holst invariant has been already studied in \cite{Racioppi:2024zva,Racioppi:2025pim}. Compatibility with data was successfully restored also in case of a sub-Planckian periodicity scale, but at the price of having a trans-Planckian mass term in front of the Holst invariant. However, in presence of torsion, an additional topological invariant can be constructed: the Nieh-Yan term (e.g. \cite{Nieh:1981, Calcagni:2009xz, Mercuri:2006um, Mercuri:2006wb, Mercuri:2009zt, Mercuri:2009zi, Date:2008rb} and refs. therein). The purpose of this work is to investigate whether and how a periodic non-minimal coupling to the Nieh-Yan term can improve the results previously obtained for natural inflation embedded in metric-affine gravity.

This article is organized as follows. In Section \ref{sec:MAG-Infl} we briefly introduce   the metric-affine framework and how to compute inflationary predictions. In Section \ref{sec:natural}, we study explicitly our new model for natural inflation in metric-affine gravity. Finally, in Section \ref{sec:conclusion} we present our conclusions.

%%%%%%%%%%%%%%%%%%%%%%%%%%%%%%%%%%%%%%%%%%%%%%%%%%%%%%%%%%%%%%%%%%%%%%%%%%%%%%%%%%%%%%%%
%%%%%%%%%%%%%%%%%%%%%%%%%%%%%%%%%%%%%%%%%%%%%%%%%%%%%%%%%%%%%%%%%%%%%%%%%%%%%%%%%%%%%%%%
\section{Metric-Affine Gravity and Inflation}\label{sec:MAG-Infl}
We begin with the Jordan frame action for a real scalar $\phi$ non-minimally coupled to metric-affine gravity:
\be 
S_{\rm J}= \int d^4x\sqrt{-g}\left\{\frac{M_P^2}{2} \left[ f(\phi){\cal R}+\tilde f(\phi)\tilde{\cal R} + 3 f_\text{NY}(\phi) \tilde{\cal T} \right]   -\frac{\partial_\mu \phi \, \partial^\mu \phi}{2} - V(\phi) \right\}, 
\label{eq:Sstart} 
\ee
where $M_P$ is the reduced Planck mass, $V(\phi)$ the inflaton potential and $f(\phi)$, $\tilde f(\phi)$ and $f_\text{NY} (\phi)$ are non-minimal coupling functions. ${\cal R}$ and $\tilde{\cal R}$ are respectively, a scalar and pseudoscalar  contraction of the curvature (the latter also known as the Holst invariant~\cite{Hojman:1980kv,Nelson:1980ph,Holst:1995pc}),
\be 
{\cal R} \equiv g^{\nu\sigma}  \tensor{\cR}{^\mu_\nu_\mu_\sigma}, \qquad  
\tilde{\cal R} \equiv  g_{\a\m} \epsilon^{\mu\nu\rho\sigma} \tensor{\cR}{^\a_\nu_\rho_\sigma},\label{eq:RRpdef}
\ee
where $\epsilon^{\mu\nu\rho\sigma}$ is the totally antisymmetric Levi-Civita tensor\footnote{The Levi-Civita tensor is defined as $\eps_{\mu\nu\rho\sigma} =\sqrt{-g} \bar{\varepsilon}_{\mu\nu\rho\sigma}$, with $ \bar{\varepsilon}_{\mu\nu\rho\sigma}$ being the Levi-Civita symbol with $ \bar{\varepsilon}_{0123}=1$. Note that the components of $\bar{\varepsilon}^{\mu\nu\rho\sigma}$ are equal to the components of ${\rm sign}(g)\bar{\varepsilon}_{\mu\nu\rho\sigma} = -\bar{\varepsilon}_{\mu\nu\rho\sigma}$.}. $\tns{\cR}{^\mu_\nu_\rho_\sigma}$ is the curvature tensor associated with the connection $\tns{\Gamma}{^\mu_\sigma_\nu}$
\be 
\tns{\cR}{^\mu_\nu_\rho_\sigma} =\partial_\rho \tns{\Gamma}{^\mu_\sigma_\nu} - \partial_\sigma \tns{\Gamma}{^\mu_\rho_\nu} +\tns{\Gamma}{^\mu_\rho_\alpha}\tns{\Gamma}{^\alpha_\sigma_\nu} -\tns{\Gamma}{^\mu_\sigma_\alpha}\tns{\Gamma}{^\alpha_\rho_\nu}  \ . \label{eq:Riemann}
\ee
We remind that in MAG, the connection  $\tns{\Gamma}{^\mu_\rho_\nu}$ is not assumed to be the Levi-Civita one, but it is computed from the corresponding equation of motion. We also remind that, if $\tns{\Gamma}{^\mu_\rho_\nu}$ is symmetric under the exchange of the lower indices, like for the Levi-Civita connection $\tns{\bar\Gamma}{^\mu_\rho_\nu}$,  $\tilde{\cal R}$ vanishes and ${\cal R}$ equals the metric Ricci scalar $R$.
Last but not least, the Nieh-Yan term\footnote{The normalization of the non-minimal coupling with the Nieh-Yan term is chosen such that $f'_\text{NY}$ and $\tilde f'$ contribute with the same numerical prefactor in eq. \eqref{eq:k(phi):general}.} $\tilde{\cal T}$ is given by  \cite{Nieh:1981, Calcagni:2009xz, Mercuri:2006um, Mercuri:2006wb, Mercuri:2009zt, Mercuri:2009zi, Date:2008rb}
\be
\tilde{\cal T} = \frac{1}{6}  \bar\nabla_\alpha \left(  \epsilon^{\a\b\c\d} T_{\b\c\d} \right)
 \label{eq:NY}
\ee
where $\bar\nabla_\alpha$ is the covariant derivative computed using the Levi-Civita connection $\tns{\bar\Gamma}{^\mu_\rho_\nu}$ and 
\be
 T^\mu_{\ \rho\sigma} = 2 \Gamma^\mu_{\ [\rho\sigma]} \, .
\label{eq:torsion}
\ee
As mentioned before, we do not consider any other term in action~\eqref{eq:Sstart} in order to keep the model as minimal as possible, with only the massless graviton and the inflaton as physical degrees of freedom  (no $\tilde{\cal R}^2$ term (e.g. \cite{Salvio:2022suk} and refs. therein)) and without terms that feature more than two derivatives (no ${\cal R}^2$ like terms (e.g. \cite{Enckell:2018hmo,Annala:2021zdt} and refs. therein)).

After some manipulations (e.g.~\cite{Langvik:2020nrs} and refs. therein), the action~\eqref{eq:Sstart} can be written in the Einstein frame as
\be 
S_{\rm E} =\int d^4x\sqrt{-g}\left[\frac{\MP^2}{2} R -\frac{1}{2}\partial_\mu \chi \, \partial^\mu \chi - U(\chi) \right],
\label{eq:SE}
\ee 
where the Einstein frame scalar potential is
\be
 U(\chi)= \frac{V(\phi(\chi))}{f(\phi(\chi))^2} \label{eq:U} \, ,
\ee
while the canonical normalized scalar $\chi$ is defined by solving
\be
  \left( \frac{d\chi}{d\phi} \right)^2= k(\phi) \, , \qquad k(\phi) = \frac{1}{f(\phi )} + \frac{6 M_P^2 \left(f'(\phi ) \tilde{f}(\phi )+f(\phi ) \left(f_\text{NY}'(\phi )-\tilde{f}'(\phi )\right)\right)^2}{f(\phi )^2 \left[f(\phi )^2+4 \tilde{f}(\phi)^2 \right]} \, ,
    \label{eq:k(phi):general}
\ee
where $'$ represents a derivative with respect to argument of the function. Now that the problem has been set in the Einstein frame, the inflationary observables can be computed following the standard equations.  First of all we define the first, second and third potential slow-roll parameters respectively as:
\bea
\epsilon_U  (\chi) &=& \frac{M_P^2}{2}\left(\frac{U'(\chi)}{U(\chi)}\right)^2 \, , \label{eq:epsilon}
\\
\eta_U  (\chi) &=& M_P^2 \frac{U''(\chi)}{U(\chi)} \, , \label{eq:eta} \\
\xi^2_U (\chi) &=& M_P^4 \frac{U'(\chi) U'''(\chi)}{U(\chi)^2} \, .
\eea
%}
The expansion of the Universe is measured in number of e-folds, which is given by
\be
N_e =  \frac{1}{M_P^2} \int_{\chi_{\textrm{end}}}^{\chi_N} {\rm d}\chi \, \frac{U(\chi)}{U'(\chi)} ,
\label{eq:Ne}
\ee
where the field value at the end of inflation is given by $\epsilon_U (\chi_{\textrm{end}}) = 1 $, while the field value $\chi_N$ at the time a given scale left the horizon is given by the corresponding $N_e$. %{\color{red} 
The tensor-to-scalar ratio $r$, the scalar spectral index $n_\textrm{s}$ and its running $\alpha_\text{s} \equiv \text{d} n_\text{s}/ \text{d} \ln k$ are respectively given by:
\bea
r  &=& 16\epsilon_U  (\chi_N) \,  , \label{eq:r} \\
n_\textrm{s}  &=& 1+2\eta_U  (\chi_N)-6\epsilon_U  (\chi_N) \, ,  \label{eq:ns} 
\\
\alpha_\textrm{s}  &=& 16 \epsilon_U (\chi_N) \eta_U (\chi_N) - 24 \epsilon_U^2 (\chi_N) - 2 \xi^2_U (\chi_N) \, . \label{eq:alphas}
\eea
%}
Finally, the amplitude of the scalar power spectrum is
\be
 A _\textrm{s} = \frac{1}{24 \pi^2 M_P^4}\frac{U(\chi_N)}{\epsilon_U  (\chi_N)}  \simeq 2.1 \times 10^{-9} \, ,
 \label{eq:As:th}
\ee
whose experimental constraint~\cite{Planck:2018jri} usually sets the energy scale of inflation.

%%%%%%%%%%%%%%%%%%%%%%%%%%%%%%%%%%%%%%%%%%%%%%%%%%%%%%%%%%%%%%%%%%%%%%%%%%%%%%%%%%%%%%%%

%%%%%%%%%%%%%%%%%%%%%%%%%%%%%%%%%%%%%%%%%%%%%%%%%%%%%%%%%%%%%%%%%%%%%%%%%%%%%%%%%%%%%%%%
\section{Applications to Natural Inflation}\label{sec:natural}
In this section we study a concrete model of inflation, defined by the natural inflaton potential
\be
 V(\phi)= \Lambda^4 \, \Omega(\phi) \label{eq:natV}
\ee
where 
\be
 \Omega(\phi) = 1+\cos\left(\frac{\phi}{M}\right) \label{eq:natfun}
\ee
and the following non-minimal couplings
\bea 
 f(\phi) &=& 1 + \xi \, \Omega(\phi) , \label{eq:f}\\
\tilde f(\phi) &=& \tilde f_0+ \tilde\xi \, \Omega(\phi) , \label{eq:ftilde}\\
f_\text{NY}(\phi) &=&  \xi_\text{NY} \, \Omega(\phi) \, .
\label{eq:fNY}
\eea
The constant term in eq. \eqref{eq:fNY} is irrelevant because $\fNY(\phi)$ contributes only via a derivative, hence we can set it to zero without loss of generality. The case with $f(\phi),\tilde f(\phi)\neq0$ and  $\fNY(\phi)=0$ has been already studied in \cite{Racioppi:2024zva,Racioppi:2025pim}, therefore such a case will be omitted in the present work. The current analysis proceeds as follows. First of all we will study only the case $f(\phi)=1$ and $\tilde f(\phi)=0$ (i.e. $\xi,\tilde \xi,\tilde f_0=0$) and assess the impact of the non-minimal coupling with the Nieh-Yan term. Then we will study the case with also the non-minimal coupling $\xi$ active.  As we will see later, such a setup will already provide viable inflationary predictions, justifying even more the absence of any coupling with the Holst invariant ($\tilde f(\phi)=0$) in the present analysis. Before proceeding, for convenience, we define 
\bea
 \Lambda &=& \delta_\Lambda M_P \, , \label{eq:delta:L} \\
 M &=& \delta_M M_P \, , \label{eq:delta:M} 
\eea
so that we can measure them in units of Planck masses. %To conclude stress that, as we work under the assumption of $\tilde f(\phi)=0$, 

\subsection{Nieh-Yan Natural Inflation}
\begin{figure}[t]
\centering
\includegraphics[width=0.6\textwidth]{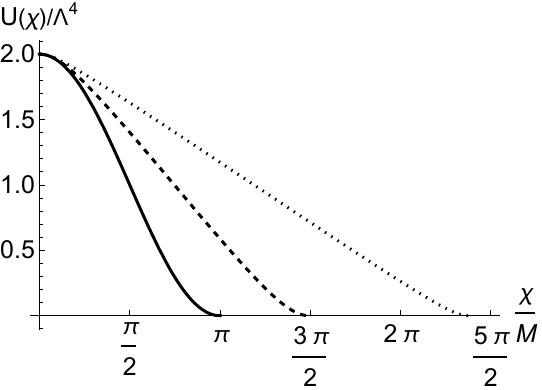}
\caption{$U(\chi)/\Lambda^4$ vs. $\chi/M$ with $M=3 M_P$ for $\xi=\tilde f =0$ and $\xi_\text{NY}=0$ (i.e. standard natural inflation) (continuous), $\xi_\text{NY}=2$ (dashed) and $\xi_\text{NY}=10$ (dotted). The graphs have been truncated in order to show only half a period. } 
\label{Fig:U:NY}
\end{figure}

\begin{figure}[tp]
\includegraphics[width=\textwidth]{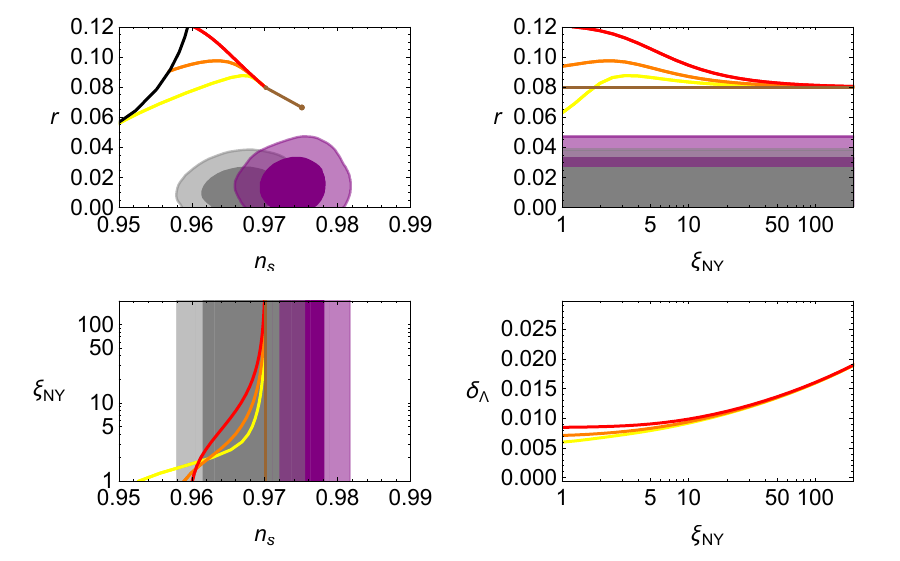}
\vspace*{-0.6cm}
\caption{$r$ vs. $n_s$ (up, left), $r$ vs. $\xi_\text{NY}$ (up, right), $\xi_\text{NY}$ vs. $n_s$ (down, left) and $\delta_\Lambda$ vs. $\xi_\text{NY}$ (down, right) for  $N_e=50$ with  $\delta_M = 5$ (yellow), $\delta_M = 7$ (orange), and $\delta_M = 10$ (red). The black line shows the predictions for standard natural inflation at $N_e=50$, while the brown line shows the predictions of linear inflation at $N_e \in [50,60]$. Contours display the $1\sigma$ and $2\sigma$ constraints from the combinations from BICEP/\emph{Keck} \cite{BICEP:2021xfz} (gray), and ACT \cite{ACT:2025tim} (purple).
} 
\label{Fig:NY:50}
\end{figure}

\begin{figure}[bp]
\vspace*{-1.3cm}
\includegraphics[width=\textwidth]{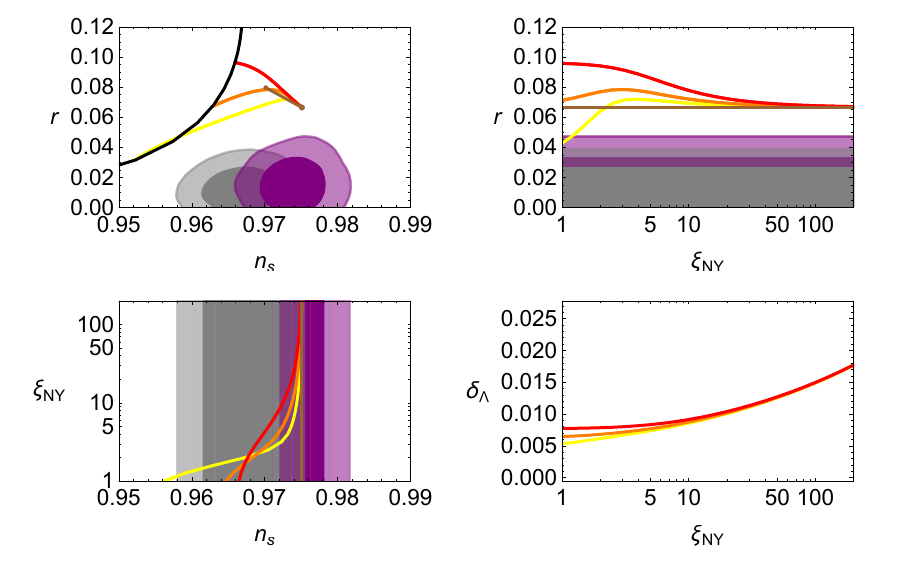}
\vspace*{-0.8cm}
\caption{Same as Fig. \ref{Fig:NY:50} but for $N_e=60$} 
\label{Fig:NY:60}
\end{figure}

In this subsection we study natural inflation in the presence of a non-minimal coupling only with the Nieh-Yan term\footnote{ The careful reader might notice that there is a degeneracy between the non-minimal couplings with the Holst and Nieh-Yan terms. Assuming $f=1$, it is possible to prove that a theory with only a non-minimal coupling $f_\text{NY}$ is equivalent to a theory with only a non-minimal coupling $\tilde f$, if the relation $f_\text{NY}(\phi) = \text{const. } \pm \frac{1}{2} \sinh ^{-1}\left(2 \tilde{f}(\phi )\right)$ is satisfied. Therefore, it would be possible to reproduce exactly the same results of \cite{Racioppi:2024zva} by only using a non-minimal coupling $f_{\text{NY}}$, but only if the ansatz in eqs. \eqref{eq:natfun}-\eqref{eq:fNY} is broken. Analogously, a more involved relation between  $f_\text{NY}(\phi)$ and $\tilde{f}(\phi )$ can be found for the case $f \neq 1$.}. Therefore, we start by applying $\xi,\tilde \xi=0$ into eq. \eqref{eq:k(phi):general} obtaining the following non minimal kinetic function 
%\be
%  \left( \frac{d\chi}{d\phi} \right)^2= k(\phi) \, , \qquad k(\phi) =1+ \frac{6 M_P^2 \left(f_{\text{NY}}'(\phi )-\tilde{f}'(\phi )\right)^2}{1+4 \tilde{f)}(\phi )^2} \, ,
%    \label{eq:k(phi):NYft0}
%\ee
%
%Using eqs. \eqref{eq:ftilde} and \eqref{eq:fNY} we obtain 
%
\be
  k(\phi) =1+ \frac{6 M_P^2 \, \xi_{\text{NY}}^2 \Big[\Omega'(\phi) \Big]^2} {1+4 {\tilde f_0}^2} \, .
    \label{eq:k(phi):NYft0}
\ee
It is well known that a kinetic function featuring a pronounced peak will induce a flattening in the canonical normalized potential. From eq. \eqref{eq:k(phi):NYft0}, we see that the larger $\tilde f_0$, the less pronounced the peak. Hence, we fix $\tilde f_0=0$, obtaining
\be
  k(\phi) =1+ 6 M_P^2 \, \xi_{\text{NY}}^2 \Big[\Omega'(\phi) \Big]^2 = 1+ 6 \left( \frac{\xi_{NY}}{ \delta_M} \right)^2 \sin ^2\left(\frac{\phi}{M}\right) \, ,
    \label{eq:k(phi):NY}
\ee
where we have used the definition of $\delta_M$ in \eqref{eq:delta:M}. Therefore, in the present scenario, the choice $\tilde f_0, \tilde \xi=0$ (i.e.  $\tilde f(\phi)=0$) favors a peaked kinetic function and therefore a flattening effect in the canonically normalized basis.
The kinetic function \eqref{eq:k(phi):NY} implies the possibility of an exact solution for $\chi(\phi)$ involving elliptic functions. However we omit such a result because it cannot be inverted exactly into $\phi(\chi)$. Nevertheless it is possible to provide numerically the graph of the function $U(\chi)$, which is given in Fig. \ref{Fig:U:NY}, where we show $U(\chi)/\Lambda^4$ vs. $\chi/M$ with $M=3 M_P$ for $\xi=\tilde f =0$ and $\xi_\text{NY}=0$ (i.e. standard natural inflation) (continuous), $\xi_\text{NY}=2$ (dashed) and $\xi_\text{NY}=10$ (dotted). We notice that by increasing $\xi_\text{NY}$, the potential gets stretched and, away from the stationary points, behaves more and more like a linear function. Such a result can be explained as follows. In the limit  $\xi _{\text{NY}} \to \infty$, away from the stationary points, the kinetic function \eqref{eq:k(phi):NY} can be approximated as
\be
  k(\phi) \simeq 6 \left( \frac{\xi_{NY}}{ \delta_M} \right)^2 \sin ^2\left(\frac{\phi}{M}\right)  \, ,
    \label{eq:k(phi):NYft0:app}
\ee
whose corresponding differential equation can be solved exactly leading to
\be
 \chi \simeq - \sqrt{6} M_P \xi _{\text{NY}} \cos \left(\frac{\phi }{M}\right)+\chi _0 \, \label{eq:chi:linear}
\ee
where $\chi_0$ is a constant of integration and we have used eq. \eqref{eq:delta:M}. Inserting \eqref{eq:chi:linear} into \eqref{eq:natV} and adjusting appropriately $\chi_0$ we obtain the linear inflation potential \cite{Andrei:2026gmu}
\be
 U(\chi) \simeq  \Lambda ^4 \left(2-\frac{\chi }{\sqrt{6} \xi _{\text{NY}} M_P }\right) \,  , \label{eq:V:linear}
\ee
in agreement with the plot in Fig. \ref{Fig:U:NY}. 
This last feature will also be confirmed by the inflationary predictions.
As mentioned before the solution $\chi(\phi)$ cannot be inverted, therefore, as common practice, we will study the inflationary phenomenology using $\phi$ as the computational variable and apply the chain-rule for derivatives. The results are
\bea
 r &=& \frac{8 \tan ^2\left(\frac{x_N}{2}\right)}{\delta _M^2+3 \xi _{\text{NY}}^2 \left[1- \cos \left(2 x_N\right)\right]} \label{eq:r:NY}\\
 n_s &=& 1 - \frac{2 \delta _M^2 \cos \left(x_N\right) \left(\cos \left(x_N\right)+1\right)+3 \sin ^2\left(x_N\right)
   \left(\delta _M^2+3 \xi _{\text{NY}}^2 \left(1-\cos \left(2 x_N\right)\right)\right)}{\left(\cos \left(x_N\right)+1\right)^2 \left(\delta _M^2+6 \xi _{\text{NY}}^2 \sin ^2\left(x_N\right)\right)^2} \qquad \label{eq:ns:NY}\\
 A_s &=& \frac{\delta _{\Lambda }^4 }{12 \pi ^2}  \Big(\delta _M^2 \csc ^2\left(x_N\right)+6 \xi_{\text{NY}}^2\Big) \Big(\cos \left(x_N\right)+1\Big)^3\label{eq:As}\\
 N_e &=& \Big[ 3 \xi _{\text{NY}}^2 \cos \left(x_N\right) \left(\cos \left(x_N\right)+2\right)-\delta _M^2 \ln \left(1-\cos
   \left(x_N\right)\right) \Big]^{x_N}_{x_\text{e}} \label{eq:Ne:NY}
\eea
where we have used the definitions in eqs. \eqref{eq:delta:L} and \eqref{eq:delta:M} and introduced $x=\phi/M$. As $\xi_\text{NY}$ contributes to the observables only via even powers, we consider only positive values for it without loss of generality.
The corresponding numerical results are given in Figs. \ref{Fig:NY:50} and \ref{Fig:NY:60}. 
In Fig. \ref{Fig:NY:50} we show $r$ vs. $n_s$ (up, left), $r$ vs. $\xi_\text{NY}$ (up, right), $\xi_\text{NY}$ vs. $n_s$ (down, left) and $\delta_\Lambda$ vs. $\xi_\text{NY}$ (down, right) for $f(\phi)=1$, $\tilde{f}(\phi)=0$ and $N_e=50$ with  $\delta_M = 5$ (yellow), $\delta_M = 7$ (orange), and $\delta_M = 10$ (red). The black line shows the predictions for standard natural inflation at $N_e=50$, while the brown line shows the predictions of linear inflation at $N_e \in [50,60]$. Contours display the $1\sigma$ and $2\sigma$ constraints from the combinations from BICEP/\emph{Keck} \cite{BICEP:2021xfz} (gray), and ACT collaborations \cite{ACT:2025tim} (purple). Fig. \ref{Fig:NY:60} is the same as Fig. \ref{Fig:NY:50} but for $N_e=60$. 
As we can see, the predictions for $r$ vs. $n_s$ are out of the observational constraints for both $N_e=50$ and $N_e=60$. Therefore, for this case, we do not compute the predictions for the running $\alpha_s$. Nevertheless, it is worth noticing that, as expected, for $\xi _{\text{NY}} \to \infty$, the predictions reach the ones of linear inflation.

\subsection{Nieh-Yan Palatini Natural Inflation} \label{subsec:PalaNY}
\begin{figure}[t]
\centering
\includegraphics[width=0.6\textwidth]{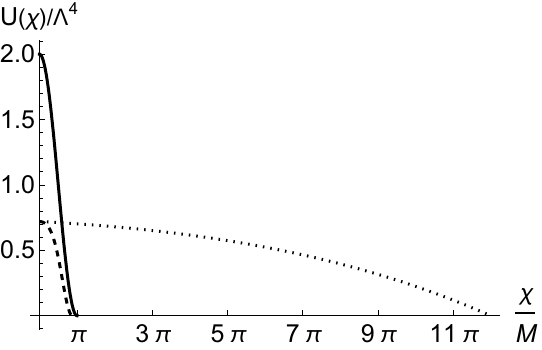}
\caption{$U(\chi)/\Lambda^4$ vs. $\chi/M$ with $\delta_M=0.01$, $\tilde f =0$  for $\xi=0$ and $\xi_\text{NY}=0$ (i.e. standard natural inflation) (continuous), $\xi=\frac{1}{3}$ and $\xi_\text{NY}=0$ (i.e. Palatini natural inflation) (dashed) and $\xi=\frac{1}{3}$ and $\xi_\text{NY}=1/10$ (dotted). The graphs have been truncated in order to show only half period. } 
\label{Fig:U:Pala:NY}
\end{figure}

We have seen in the previous subsection that the non-minimal coupling with the Nieh-Yan term drives the inflationary predictions towards the ones of linear inflation. We check now whether the inclusion of a non-minimal coupling with the Ricci scalar drives the inflationary predictions into the regions allowed by data. In this case the kinetic function \eqref{eq:k(phi):general} becomes
\be
 k(\phi) = \frac{1}{1+\xi  \left(1+\cos \left(\frac{\phi }{M}\right)\right)} + \frac{\frac{6 \xi _{\text{NY}}^2 }{\delta _M^2} \sin ^2\left(\frac{\phi }{M}\right)}{\left(1+\xi  \left(1+\cos \left(\frac{\phi }{M}\right)\right)\right)^2} \, , \label{eq:k:Pala:NY}
\ee
It has been proven \cite{Racioppi:2024zva,Racioppi:2025pim}, that a non-minimal coupling with pure Palatini gravity (i.e. $\tilde f(\phi)=f_\text{NY}(\phi) =0$) can rescue natural inflation for $\delta_M > 1$ and that the most favored predictions are obtained when $\xi \simeq \frac{1}{3}$. Therefore from now on we adopt this choice and keep only $\xi_\text{NY}$ as free parameter. Moreover, we also make the additional assumption of $\delta_M=0.01$ in order to check if our new scenario can rescue natural inflation when the periodicity scale is sub-Planckian. Before proceeding with the inflationary analysis, we show in Fig. \ref{Fig:U:Pala:NY} $U(\chi)/\Lambda^4$ vs. $\chi/M$ with $\delta_M=0.01$, $\tilde f =0$  for $\xi=0$ and $\xi_\text{NY}=0$ (i.e. standard natural inflation) (continuous), $\xi=\frac{1}{3}$ and $\xi_\text{NY}=0$ (i.e. Palatini natural inflation) (dashed) and $\xi=\frac{1}{3}$ and $\xi_\text{NY}=1/10$ (dotted). We can appreciate that the non-minimal coupling $\xi$ lowers the height of the maximum of the potential and then the non-minimal coupling $\xi_\text{NY}$ stretches the potential creating an enlarged flat region. In such a case the corresponding inflationary predictions are
\bea
 r &=& \frac{8 \left(\cos \left(x_N\right)-2\right)^2 \tan ^2\left(\frac{x_N}{2}\right)}{3 \left(\delta _M^2 \left(\cos
   \left(x_N\right)+4\right)+9 \xi _{\text{NY}}^2 \left(1-\cos \left(2 x_N\right)\right)\right)} \, , \label{eq:r:Pala:NY}\\
 n_s &=& 1 -\frac{\left(\cos \left(x_N\right)-2\right)^2 \tan ^2\left(\frac{x_N}{2}\right)}{\delta _M^2 \cos \left(x_N\right)+4 \delta _M^2-9 \xi
   _{\text{NY}}^2 \cos \left(2 x_N\right)+9 \xi _{\text{NY}}^2} + \label{eq:ns:Pala:NY} \\
      && \Bigg[18 \cos \left(3 x_N\right) \left(2 \delta _M^2-3 \xi _{\text{NY}}^2\right)+4 \cos \left(x_N\right) \left(9 \xi
   _{\text{NY}}^2-113 \delta _M^2\right)+\nn\\
   &&8 \cos \left(2 x_N\right) \left(37 \delta _M^2+144 \xi _{\text{NY}}^2\right)-\cos\left(4 x_N\right) \left(\delta _M^2+288 \xi _{\text{NY}}^2\right)-9 \left(31 \delta _M^2+96 \xi _{\text{NY}}^2\right)+\nn\\
   &&+18 \xi
   _{\text{NY}}^2 \cos \left(5 x_N\right)\Bigg] \Bigg/ \Bigg[ 24 \left(\cos \left(x_N\right)+1\right) \left(\delta _M^2 \left(\cos    \left(x_N\right)+4\right)+18 \xi _{\text{NY}}^2 \sin ^2\left(x_N\right)\right)^2 \Bigg] \nn\\
 A_s &=& \frac{9 \delta _{\Lambda }^4 \left(\cos \left(x_N\right)+1\right)^3 \left(\delta _M^2 \left(\cos \left(x_N\right)+4\right) \csc ^2\left(x_N\right)+18 \xi _{\text{NY}}^2\right)}{4 \pi ^2 \left(\cos \left(x_N\right)-2\right)^2 \left(\cos \left(x_N\right)+4\right)^2} \, , \label{eq:As:Pala:NY} \\
 N_e &=& 3 \Bigg[\left(\delta _M^2-9 \xi _{\text{NY}}^2\right) \ln \left(2-\cos
   \left(x_N\right)\right)-2 \delta _M^2 \ln \left(\sin \left(\frac{x_N}{2}\right)\right) + \nn\\
  && -9 \xi _{\text{NY}}^2 \ln \left(\cos \left(x_N\right)+4\right) \Bigg]^{x_N}_{x_\text{e}} \, . \label{eq:Ne:Pala:NY}
\eea
In this case, it is also needed to check the running of the spectral index, which is
\bea
\alpha_s &=& - \Bigg[\sin ^2\left(\frac{x_N}{2}\right) \cos ^4\left(\frac{x_N}{2}\right) \left(\cos \left(x_N\right)-2\right) \left(\cos
   \left(x_N\right)+4\right) \times \nn\\
   && \quad \times \Big(6 \cos \left(x_N\right) \left(-1007 \delta _M^4+4302 \delta _M^2 \xi _{\text{NY}}^2+2916 \xi _{\text{NY}}^4\right)+\nn\\
   &&\qquad + 48 \cos \left(2 x_N\right) \left(-28 \delta _M^4+909 \delta _M^2 \xi _{\text{NY}}^2+1458 \xi_{\text{NY}}^4\right) +\nn\\
   &&\qquad -9 \cos \left(3 x_N\right) \left(29 \delta _M^4-342 \delta _M^2 \xi _{\text{NY}}^2+2916 \xi _{\text{NY}}^4\right) +\nn\\
   &&\qquad -4 \cos \left(4 x_N\right) \left(7    \delta _M^4+468 \delta _M^2 \xi _{\text{NY}}^2+4374 \xi _{\text{NY}}^4\right) + \nn\\
   &&\qquad +\cos \left(5 x_N\right) \left(-\delta _M^4-90 \delta _M^2 \xi
   _{\text{NY}}^2+8748 \xi _{\text{NY}}^4\right) -36 \left(209 \delta _M^4+360 \delta _M^2 \xi _{\text{NY}}^2+1458 \xi
   _{\text{NY}}^4\right)\Big) \Bigg] \times \nn\\
   && \times \Bigg[9 \left(\cos \left(x_N\right)+1\right)^4 \Big(\delta _M^2 \left(\cos \left(x_N\right)+4\right)+18 \xi
   _{\text{NY}}^2 \sin ^2\left(x_N\right)\Big)^4 \Bigg]^{-1}
    \, . \label{eq:alphasPala:NY}
\eea

\begin{figure}[t]
\includegraphics[width=\textwidth]{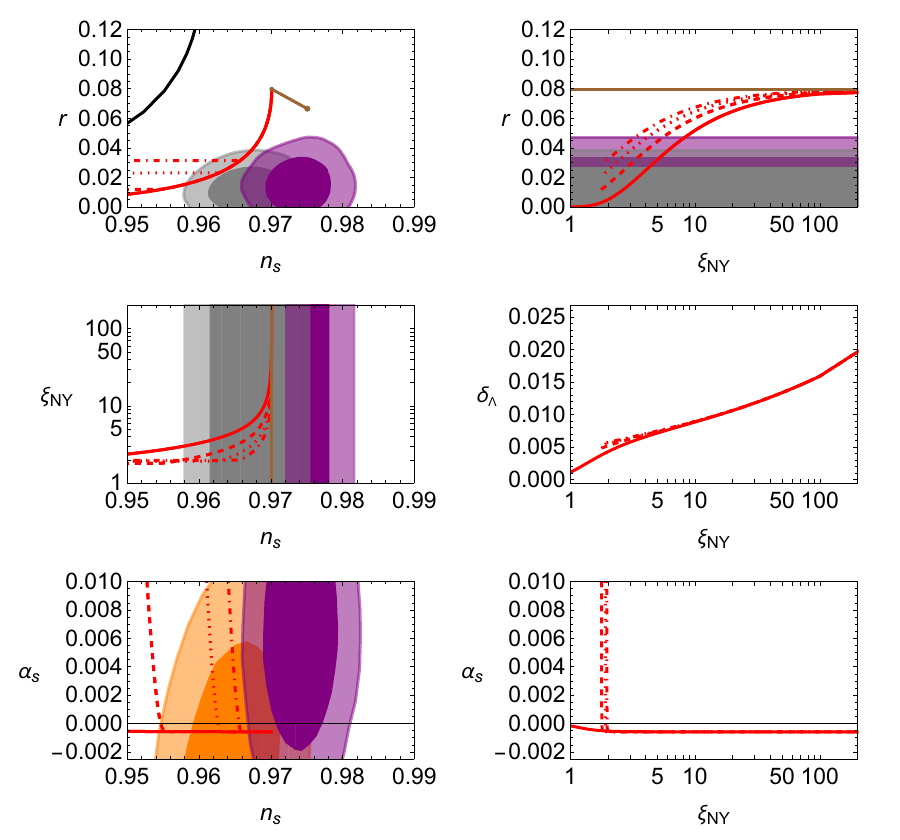} 
\caption{$r$ vs. $n_s$ (up, left), $r$ vs. $\xi_\text{NY}$ (up, right), $\xi_\text{NY}$ vs. $n_s$ (center, left),  $\delta_\Lambda$ vs. $\xi_\text{NY}$ (center, right), $\alpha_s$ vs. $n_s$ (down, left),  $\alpha_s$ vs. $\xi_\text{NY}$ (down, right) with $\delta_M = 0.01$, $\tilde{f}(\phi)=0$ and $N_e=50$ (red) for $\xi=\frac{1}{2}$ (continuous), $\xi=\frac{1}{3}$ (dashed),  $\xi=\frac{1}{4}$ (dotted) and  $\xi=\frac{1}{5}$ (dot-dashed). The black line represents the predictions of standard natural inflation. The brown line shows the predictions of linear inflation at $N_e \in [50,60]$. Contours display the $1\sigma$ and $2\sigma$ constraints from the combinations from Planck (orange) \cite{Planck:2018vyg}, BICEP/\emph{Keck} \cite{BICEP:2021xfz} (gray), and ACT collaborations \cite{ACT:2025tim} (purple).} \label{fig:Pala:NY:50}
\end{figure}
\begin{figure}[t]
\includegraphics[width=\textwidth]{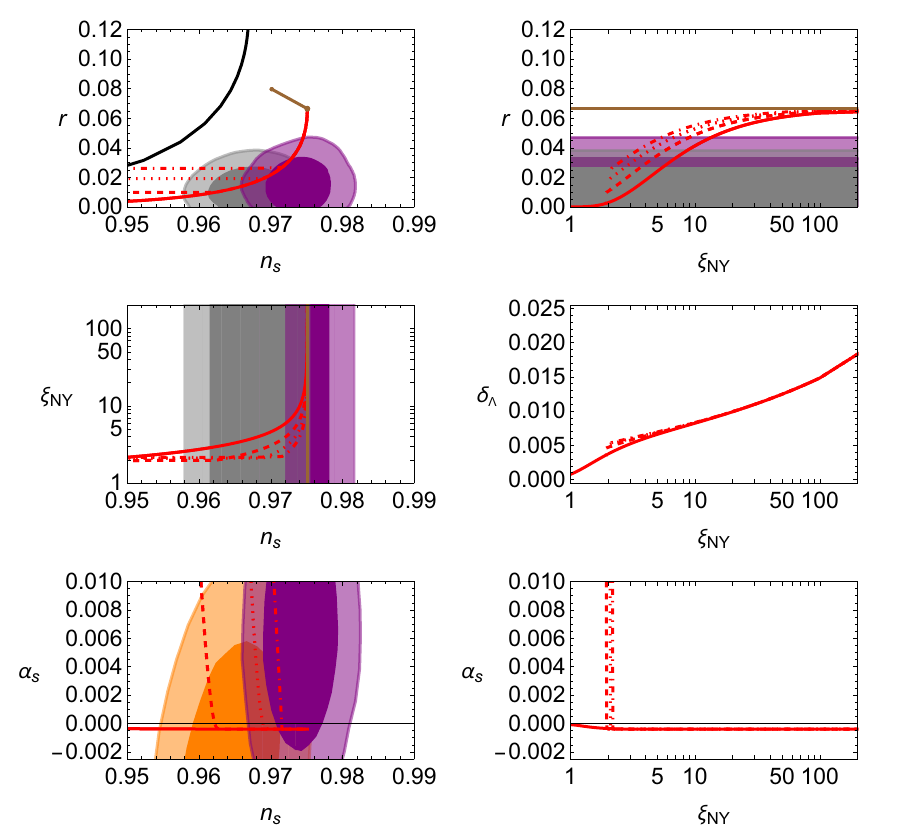} 
\caption{Same as Fig. \ref{fig:Pala:NY:50} but for $N_e=60$} \label{fig:Pala:NY:60}
\end{figure}
Again, as $\xi_\text{NY}$ contributes to the observables only via even powers, we consider only positive values for it without loss of generality. {
The corresponding results are given in Figs. \ref{fig:Pala:NY:50} and \ref{fig:Pala:NY:60}. Fig.  \ref{fig:Pala:NY:50}  shows $r$ vs. $n_s$ (up, left), $r$ vs. $\xi_\text{NY}$ (up, right), $\xi_\text{NY}$ vs. $n_s$ (center, left),  $\delta_\Lambda$ vs. $\xi_\text{NY}$ (center, right), $\alpha_s$ vs. $n_s$ (down, left),  $\alpha_s$ vs. $\xi_\text{NY}$ (down, right) with $\delta_M = 0.01$, $\tilde{f}(\phi)=0$ and $N_e=50$ (red) for $\xi=\frac{1}{2}$ (continuous), $\xi=\frac{1}{3}$ (dashed),  $\xi=\frac{1}{4}$ (dotted) and  $\xi=\frac{1}{5}$ (dot-dashed). The black line represents the predictions of standard natural inflation. The brown line shows the predictions of linear inflation at $N_e \in [50,60]$. Contours display the $1\sigma$ and $2\sigma$ constraints from the combinations from Planck (orange) \cite{Planck:2018vyg}, BICEP/\emph{Keck} \cite{BICEP:2021xfz} (gray), and ACT collaborations\footnote{It is known that running can slightly affect the exact position of the $r$ vs. $n_s$ contours. However, for simplicity, such contours are drawn assuming negligible running i.e. $\alpha_s \simeq 0$. This does not sensibly affect our results because most of the allowed predictions exhibit $|\alpha_s| \lesssim 0.001$.} \cite{ACT:2025tim} (purple).
Fig. \ref{fig:Pala:NY:60} is the same as \ref{fig:Pala:NY:50} but for $N_e=60$. First of all we notice that $\xi_\text{NY}$ exhibits a non-zero lower bound (with the exact value slightly depending on $N_e$ and $\xi$) except for the $\xi=\frac{1}{2}$ case. This is due to the fact that natural inflation suffers from the $\eta$-problem at small values of $M$ and the same could happen in our scenario as well.
This can be easily seen by evaluating the second potential slow-roll parameter\footnote{In this article we consider only $\xi \leq \frac{1}{2}$ in order to avoid the issue of the maximum of the potential being displaced from $\phi=0$ to $\phi=M \arccos\left(\frac{1-\xi }{\xi }\right)$ and the consequent loss of the reflection symmetry around the maximum \cite{Racioppi:2024zva}. Nevertheless the same argument about the $\eta$-problem can be extended to the case of $\xi > \frac{1}{2}$. Moreover the limit shown in eqs. \eqref{eq:rN:Pala:NY:limit}-\eqref{eq:asN:Pala:NY:limit} is valid also for $\xi > \frac{1}{2}$.} (see eq.~\eqref{eq:eta}) at $\phi \ll M_P$:
\begin{eqnarray}
    \eta_U(\phi\ll M_P) &\simeq & \frac{(2 \xi -1) }{2 \delta _M^2} \left(1-\frac{12 \xi _{\text{NY}}^2}{(2 \xi +1) \delta _M^2} \left(\frac{\phi}{M}\right)^2 \right) \, .
    %\nn\\  &\simeq & -\frac{5000}{3} \left(1-720000000 \xi _{\text{NY}}^2 \frac{ \phi ^2}{M_P^2}\right) \, , 
    \label{eq:eta:0_exp:full} 
\end{eqnarray}
%
%where in the last line we have inserted the actual values of $\xi=\frac{1}{3}$ and $\delta_M=0.01$.
It can be proven that the smaller $M$, the smaller is $\phi_N$ (the field value corresponding to $N_e$). Therefore, when $\xi _{\text{NY}} \to 0$, we have $|\eta_U(0)| \sim  \frac{(2 \xi -1) }{2 \delta _M^2}$, and slow-roll is preserved only within a tiny interval around $\xi=\frac{1}{2}$: $\xi \in \frac{1}{2} \left[1-2 \delta _M^2,1+2 \delta_M^2\right]$. Indeed, when  $\xi=\frac{1}{3},\frac{1}{4},\frac{1}{5}$, the lines at $\xi _{\text{NY}}=0$ are not visible in Figs.~\ref{fig:Pala:NY:50} and \ref{fig:Pala:NY:60}. On the other hand, with $\xi _{\text{NY}}$ increasing it is possible to restore the validity of the slow-roll approximation and even get predictions in the allowed region of the combination from the BICEP/\emph{Keck} collaboration~\cite{BICEP:2021xfz} and even the allowed region of the combination from the ACT collaboration \cite{ACT:2025tim}. However, since $|\eta_U(0)| \gg 1$ in order to get $n_s$ in the allowed region, $|\eta_U|$ needs to decrease very fast by increasing $\xi _{\text{NY}}$, implying a very big running of the spectral index $\alpha_s$ which is ruled out by Planck legacy data \cite{Planck:2018vyg}. All of this happens in a very small range of $\xi _{\text{NY}}$. 
Then by $\xi_\text{NY}$ increasing, the running of $n_s$ is reduced and $r$ and $n_s$ are increasing and entering the allowed region of only the combination from BICEP/\emph{Keck} or also the one of the combination from ACT, according to value of $N_e$, until the linear limit is activated and the predictions exit again both allowed regions. This last feature has an explanation quite similar to the previous case. It can be proven that  if $\xi_\text{NY} \to \infty$, then $x_N \to \pi$ implying $ (1+ \cos x_N) \to 0$ and therefore the kinetic function behaves like the one in eq. \eqref{eq:k(phi):NY}. Then the same argument of the previous case applies here. Moreover we also notice that, as soon as the $\eta$-problem is solved, all the predictions for $r$ vs. $n_s$ and $\alpha_s$ vs. $n_s$ overlap with each other. The motivation is the following. If the periodicity scale is very much smaller than $M_P$, like in our case of interest $\delta_M=0.01$, and the inflationary regime is enough displaced from $\phi=0$, then the second term of the kinetic function \eqref{eq:k:Pala:NY} is completely dominant over the first one, allowing the approximation
\be
 k(\phi) \simeq \frac{\frac{6 \xi _{\text{NY}}^2 }{\delta _M^2} \sin ^2\left(\frac{\phi }{M}\right)}{\left(1+\xi  \left(1+\cos \left(\frac{\phi }{M}\right)\right)\right)^2} \, , \label{eq:k:Pala:NY:app}
\ee
which leads to an analytical solution of eq. \eqref{eq:k(phi):general}
\be
 \chi \simeq \sqrt{6} M_P \frac{ \xi _{\text{NY}} }{\xi } \ln\left(\frac{2 \xi +1}{1+\xi  \left(1+\cos \left(\frac{\phi }{M}\right)\right)}\right) \label{eq:chi:app:PalaNY} \, .
\ee
Applying eq. \eqref{eq:chi:app:PalaNY} to eq. \eqref{eq:Ne:Pala:NY}, we can approximate $N_e$ as
\be
 N_e \simeq \frac{3 \xi _{\text{NY}}^2 }{\xi ^2} \ln \left(\frac{e^{\frac{\sqrt{\frac{2}{3}} \chi  \xi }{M_P \xi _{\text{NY}}}}}{(2
   \xi +1) \left(2 e^{\frac{\chi  \xi }{\sqrt{6} M_P \xi _{\text{NY}}}}-(2 \xi+1)\right)}\right) \, . \label{eq:Ne:app:PalaNY}
\ee
Applying eqs. \eqref{eq:chi:app:PalaNY} and \eqref{eq:Ne:app:PalaNY} to eqs. \eqref{eq:r:Pala:NY}-\eqref{eq:alphasPala:NY} we obtain
\bea
 r &\simeq & \frac{2}{3} \frac{\xi^2}{\xi_\text{NY}^2} \left(\coth \left(\frac{\xi^2}{\xi_\text{NY}^2} \frac{N_e}{6}\right)-2 \sqrt{e^{\frac{\xi^2}{\xi_\text{NY}^2} \frac{N_e}{3}}-1} \text{csch}\left(\frac{\xi^2}{\xi_\text{NY}^2} \frac{N_e}{6}\right)+3\right) \, , \label{eq:rN:Pala:NY:limit}\\
 n_s &\simeq &  1-\frac{1}{12} \frac{\xi^2}{\xi_\text{NY}^2} \left(3 \coth \left(\frac{\xi^2}{\xi_\text{NY}^2} \frac{N_e}{6}\right)+5\right) \, , \\
 A_s &\simeq &  \frac{\delta _{\Lambda }^4 }{2 \pi ^2 } \frac{\xi _{\text{NY}}^2}{\xi^3}  e^{\frac{\xi ^2}{ \xi _{\text{NY}}^2} \frac{N_e}{6}} \left(e^{\frac{\xi ^2}{\xi _{\text{NY}}^2} \frac{N_e}{3}}-1\right)^{3/2} \, , \\
 \alpha_s &\simeq &  -\frac{1}{24} \frac{\xi^4}{\xi_\text{NY}^4} \text{csch}^2\left(\frac{\xi^2}{\xi_\text{NY}^2} \frac{N_e}{6}\right) \label{eq:asN:Pala:NY:limit} \, ,
\eea
where we can see that $r$, $n_s$ and $\alpha_s$ depend only on the ratio $\frac{\xi _{\text{NY}}}{\xi}$, in agreement with what shown in Figs. \ref{fig:Pala:NY:50} and \ref{fig:Pala:NY:60}. Additionally we also note that all the observables in eqs.  \eqref{eq:rN:Pala:NY:limit}-\eqref{eq:asN:Pala:NY:limit} do not depend on $\delta_M$, meaning that the numerical results are valid for any $\delta_M \ll 1$ as long as the $\eta$-problem is avoided.
To conclude, it is worth noticing that agreement with data can be achieved with $6 \lesssim \frac{\xi _{\text{NY}}}{\xi} \lesssim 30$ and $\delta_\Lambda \sim 0.01$ when $\xi$ and $\xi _\text{NY}$ are so that the $\eta$-problem is avoided.

}
%%%%%%%%%%%%%%%%%%%%%%%%%%%%%%%%%%%%%%%%%%%%%%%%%%%%%%%%%%%%%%%%%%%%%%%%%%%%%%%%%%%%%%%%
\section{Conclusions}\label{sec:conclusion}

In this paper, we constructed and analyzed a new metric-affine realization of natural inflation. We focused on a model containing only the graviton and the PNGB inflaton in the particle content. The setup features two non-minimal couplings: one to the Ricci scalar, parametrized by $\xi$, and another to the Nieh-Yan term, parametrized by $\xi_{\text{NY}}$.

{ Since previous non-minimally coupled realizations of natural inflation can already reconcile the model with observations, if a trans-Planckian periodicity scale $M$ is considered, we focused our study on $\delta_M \ll 1$, using the reference value $\delta_M = 0.01$ where needed. We investigated the following representative cases, $\xi = 0$ (corresponding to a scenario in which the non-minimality is due solely to the Nieh-Yan term), $\frac{1}{5}, \frac{1}{4}, \frac{1}{3}, \frac{1}{2}$. Consequently, $\xi_{\text{NY}}$ remained the only free parameter in our analysis.}

We performed a numerical study of the Einstein-frame potential as a function of the canonically normalized inflaton $\chi$. For very large $\xi_{\mathrm{NY}}$, the potential develops an approximately linear behaviour away from its stationary points in both cases considered. This feature is also reflected in the inflationary predictions. However, the $\xi = 0$ scenario was not compatible with observational constraints for the full range considered for the number of $e$-folds $N_e$ also for trans-Planckian values of $M$. { In contrast, the $\xi \neq 0$ cases achieved agreement with observational data. For sufficiently large $\xi_{\mathrm{NY}}$, the $\eta$-problem (induced by the choice $\delta_M \ll 1$) was solved and the predictions for $r$, $n_s$, $\alpha_s$ were numerically depending only on the ratio $\frac{\xi_\text{NY}}{\xi}$ but not on the exact value of $\delta_M$. Depending on the values of $\frac{\xi_\text{NY}}{\xi}$, the predictions were compatible with the combination from BICEP/\emph{Keck}, ACT, or both.

Even though the predictions within observational constraints are obtained without invoking initial trans-Planckian mass scales, the issue of trans-Planckian field fluctuations is still present in the canonical base, as visible from Fig. \ref{Fig:U:Pala:NY}. This is due to the effective strong coupling regime, generated by $\delta_M \ll 1$ in eq. \eqref{eq:k:Pala:NY:app}, which induces a large kinetic stretching and therefore a trans-Planckian effective periodicity scale in the canonical base.
This is a well known shortcoming of natural inflation scenarios, whose solution usually requires additional structure, such as alignment mechanisms, kinetic mixing effects, or multiple axion fields (see e.g. \cite{Pajer:2013fsa} and refs. therein). Such a completion is beyond the scope of our analysis and we leave it for future studies.

Finally we stress that compatibility with data requires only moderately large non-minimal couplings satisfying $6 \lesssim \frac{\xi _{\text{NY}}}{\xi} \lesssim 30$.
}
%%%%%%%%%%%%%%%%%%%%%%%%%%%%%%%%%%%%%%%%%%%%%%%%%%%%%%%%%%%%%%%%%%%%%%%%%%%%%%%%%%%%%%%%
%%%%%%%%%%%%%%%%%%%%%%%%%%%%%%%%%%%%%%%%%%%%%%%%%%%%%%%%%%%%%%%%%%%%%%%%%%%%%%%%%%%%%%%%
\acknowledgments
%%%%%%%%%%%%%%%%%%%%%%%%%%%%%%%%%%%%%%%%%%%%%%%%%%%%%%%%%%%%%%%%%%%%%%%%%%%%%%%%%%%%%%%%
This work was supported by the Estonian Research Council grants PRG1677, TARISTU24-TK10, TARISTU24-TK3, and by the CoE program TK202 ``Foundations of the Universe''.  This article is based upon work from COST Actions COSMIC WISPers CA21106 and CosmoVerse CA21136, supported by COST (European Cooperation in Science and Technology).

\bibliographystyle{JHEP}
\bibliography{references}

\end{document}